\documentclass[11pt]{article}
\usepackage{latexsym}
\usepackage{amssymb}
\usepackage[dvips]{graphicx}
\usepackage{epsfig}
\usepackage{rotating}

\usepackage{amsfonts}
\usepackage{amsmath}
\begin{document}

\begin{titlepage}
\begin{flushright}
NITheP-09-12\\
ICMPA-MPA/2009/21\\
\end{flushright}

\begin{center}

{\Large\bf $q$-graded Heisenberg algebras
and deformed supersymmetries}

Joseph Ben Geloun$^{a,b,c,*}$ and Mahouton Norbert Hounkonnou$^{b,\dag}$

$^{a}${\em National Institute for Theoretical Physics (NITheP)}\\
{\em Private Bag X1, Matieland 7602, South Africa}\\
$^{b}${\em International Chair in Mathematical Physics
and Applications}\\
{\em (ICMPA--UNESCO Chair), 072 B.P. 50  Cotonou, Republic of Benin}\\
$^{c}${\em D\'epartement de Math\'ematiques et Informatique}\\
{\em  Facult\'e des Sciences et Techniques, Universit\'e Cheikh Anta Diop, Senegal}

E-mails:  $^{*}$bengeloun@sun.ac.za,\quad $^{\dag}$norbert.hounkonnou@cimpa.uac.bj

\begin{abstract}
The notion of $q$-grading on the enveloping algebra generated by
products of $q$-deformed Heisenberg algebras is introduced for $q$ complex number
in the unit disc. Within this formulation, we consider the extension of the notion
of supersymmetry in the enveloping algebra.
We recover the ordinary $\mathbb{Z}_2$ grading
or Grassmann parity for associative superalgebra,
and  a modified version of the usual supersymmetry.
As a specific problem, we focus on the interesting limit $q\to -1$ for
which the Arik and Coon deformation of the Heisenberg algebra
allows to map fermionic  modes to bosonic ones in a modified sense. Different algebraic
consequences are discussed.
\end{abstract}

\today
\end{center}
Pacs numbers: 03.65.Fd, 03.65.-w, 11.30.Pb\\
Keys words: $q$-oscillator algebras, associative superalgebra, supersymmetry.

\end{titlepage}

\section{Introduction}
\label{sect:intro}

Deformations of the bosonic Heisenberg algebra by parameters have known successful
achievements in mathematical physics \cite{ar}-\cite{nk3}
and in nonlinear physics (e.g. nonlinear quantum optics) \cite{manko}-\cite{qsd}.
One of the simplest deformation of the bosonic algebra,
a one parameter $q$-deformation, was introduced by
Arik and Coon \cite{ar} and is defined by
\begin{eqnarray}
aa^{\dag}-qa^{\dag}a=\mathbb{I}, \quad 0 < q \leq 1.
\end{eqnarray}
Clearly, one recovers the ordinary Fock algebra of the harmonic
oscillator at the limit $q \to 1$, with then
$[a,a^\dagger]=\mathbb{I}$.

The most of studies pertaining to such  deformations
are made with the parameter $q\in ]0,1]$.
However, in \cite{ernst}, a study was performed even for complex
values of $q$. Concerning the issue of convergence,
infinite products and deformed
exponential series require at least that the modulus $|q|\leq 1$.
This leads to the consideration that $q \in \mathbb{D^*}$, with
$\mathbb{D^*}$ the unit complex disc but the zero.
Keeping in mind these last remarks, nothing prevents to perform the following
limit
\begin{eqnarray}
\lim_{q\to -1} aa^{\dag}-qa^{\dag}a= aa^{\dag}+a^{\dag}a= \{a,a^\dag\}
=\mathbb{I},
\label{ferm}
\end{eqnarray}
reminiscent of a fermionic algebra  \cite{haya}-\cite{bag}.
It then raises many natural questions.
Is it possible to understand the generators associated to this
limit as fermions ? Then, in the case of a positive answer,
is there a mapping from the bosonic operators
(defined for $q=1$) to fermionic ones (defined for $q=-1$),
i.e. a kind of supersymmetry ?

Recent years, many investigations on $q$-deformed algebras and supersymmetry
have been undertaken dealing with  $q$-deformed supersymmetric
factorization \cite{ioffe} or differential representation,
intertwining properties and coherent states \cite{bag}
(and more references therein).
Nevertheless, as far as we can establish, none of them focuses on the
complete study of the product of these deformed algebras
for different parameters $q$. So doing, one will immediately generate
a full deformed universal algebra of all different deformed generators
acting on a unique representation Hilbert space.
We propose to investigate how the notions of $\mathbb{Z}_2$ Grassmann
grading and supersymmetry can be extended to this multi-deformed enveloping algebra.

In this paper, we introduce the notion of $q$-deformed grading
on the enveloping algebra generated by all products in
different deformed Heisenberg algebras. This notion generalizes
the ordinary Grassmann grading and, moreover, by defining a generalized
$q$-graded bracket, one is able to recover, in
each subalgebra, the correct structure for bosonic, fermionic,
$\mathbb{Z}_2$ graded and basic deformed bosonic algebras.
The extension of Grassmann parity affords us to understand
ordinary associative superalgebras and their $\mathbb{Z}_2$ graded structure
(the  usual framework of supersymmetry) as limit algebras
when the parameter $q\to \pm 1$. We then determine the modified supersymmetric
Hamiltonian and its deformed supercharges mapping some deformed fermions on
deformed bosons.

The paper's outline is as follows. The following section is
dedicated to the definition and basic properties of the
$q$-deformation of the Heisenberg algebra in the sense of Arik and
Coon, for complex parameter $q$, and its representation. The limit
$q\to -1$ is clarified. Afterwards, Section 3 addresses the
algebraic settlement of the deformed structure producing the general
$q$-deformed grading. The particular case of ordinary Grassmann
parity is discussed. Section 4 investigates the extended notion of
supersymmetry on the enveloping algebra. The specific limit $q\to
\pm 1$, producing a modified version of the ordinary supersymmetry,
is also discussed. The paper ends by some remarks in Section 5 and
an appendix provides useful identities and illustrations.

\section{Complex $q$-deformed Heisenberg algebras}
\label{sect2}

Let us consider the Arik and Coon deformation of the Heisenberg algebra \cite{ar}
\begin{eqnarray}
a_qa_q^\dag -q a_q^\dag a_q =\mathbb{I},
\label{qdef0}
\end{eqnarray}
with parameter a complex number $q$. If we regard $a^\dag_q$ as the adjoint
of $a_q$, it follows that, by Hermitian conjugation of (\ref{qdef0}),
$a_qa_q^\dag -\bar q a_q^\dag a_q =\mathbb{I}$.
By simple substraction of these equations,
one ends with $\bar q=q$ from the positivity of $a_q^\dag a_q$. Hence, $q$ should be
a real parameter. However, introducing a new operator $a_q^\natural$, let us
reconsider the same kind of deformed structure, namely
\begin{eqnarray}
a_qa_q^\natural - q a_q^\natural a_q =\mathbb{I},
\label{qdef}
\end{eqnarray}
and relax the previous condition of adjoint property between $a_q$ and $a_q^\natural$.
Then, nothing can be said, {\it a priori}, on the parameter $q$.
We will place ourself in this general situation such that\footnote{
In fact, this condition may not be imposed because
nothing prevents to do an extension $q\in \mathbb{C}$, $\varphi_q\in [0,2\pi[$,
$q\neq 0$.
All the following main equations are again valid outside the unit disc.
Only the notion of convergence of functions series and infinite products
involved in deformed special function theory has to be reconsidered. We are not
dealing with these ideas here, but we want, as much as possible, to have
a theory with interesting properties for the theoretician community.}
\begin{eqnarray}
&&q\in \mathbb{D}^* = \{z\in \mathbb{C}, \;|z|\leq 1,\; z\neq 0\},\cr
&& q = r_q e^{i\varphi_q}, \;\; r_q\in ]0,1],\;\; \varphi_q \in [0,2\pi[.
\end{eqnarray}
In order to define the power function of $q$, namely $q^x$, one uses the
complex form of the exponential function $e^{x\, {\rm Ln}q }$, where
${\rm Ln}(\cdot)$ stands for the principal branch of complex logarithm.

A realization of the algebra (\ref{qdef}) is also well known.
To construct it, one starts with the ordinary bosonic operators $a$ and $a^\dag$,
fulfilling $[a,a^\dagger]=\mathbb{I}$ with the number  $N=a^\dag a$,
generating the ordinary number operator in the Fock Hilbert space
${\mathcal H}={\rm span}\{|n\rangle= (1/\sqrt{n!})(a^\dagger)^n|0\rangle\}$. Then we define
\begin{eqnarray}
a_q |n\rangle=\sqrt{\frac{[N+1]_q}{N+1}}a|n\rangle,\quad
a_q^\natural |n\rangle=\sqrt{\frac{[N]_q}{N}}a^\dagger|n\rangle,\quad
 [N]_q = \frac{1-q^N}{1-q},
\end{eqnarray}
where one refers to $[N]_q$  as the $q$-basic number of the
theory. Note that  $[N]_q$ is not necessarily  self-adjoint.
Indeed, $([N]_q)^\dag=[N]_{\bar q}$, which is not $[N]_q$ unless $q$ is real.
The adjoint of the operator $a_q$ can be expressed as
\begin{eqnarray}
a_q^\dag =  \sqrt{\frac{[N]_{\bar q}}{N}}a^\dagger
\label{adjoi}
\end{eqnarray}
from which it appears possible to define naturally a self-adjoint
deformed number operator as $\{N\}_{\bar q,q} :=a_q^\dag a_q =\sqrt{{[N]_{\bar q}[N]_q}}$.
From (\ref{adjoi}), a relation between $a_q^\dag$ and $a_q^\natural$
can be inferred:
\begin{eqnarray}
a_{\bar q}^\natural=a_q^\dag\quad \Leftrightarrow\quad
a_{q}^\natural=\overline{(a_q^\dag)}=a_{\bar q}^\dag.
\label{relat}
\end{eqnarray}
We are then in position to define properly the unary operation $\natural$
which is the adjoint operation composed with the complex conjugation.
The operator $a_{q}^\natural=\overline{(a_q^\dag)}=a_q^t$,
viewed as a matrix, can be understood as the transpose of $a_q$.
Moreover, it can be checked that $(a_{q}^\natural)^\natural= a_q$, therefore $\natural$
is an involution; we also have $(a_q^\natural a_q )^\natural= a_q^\natural a_q $.
For a real parameter $q$, the definitions of $a^\dag_q$ and $a_{q}^\natural$
coincide.

Let us briefly mention the limit $q\to 0$. The corresponding basic
number $[N]_0$ proves to be the constant operator $\mathbb{I}$. This
implies that $a_0$ and $a_0^\dag$ are mutually inverse in the Fock space
without the vacuum $|0\rangle$. As a result of the triviality of the $(q=0)$-commutator,
the $(q=0)$-deformed algebra is again a Lie algebra. Then the enveloping
algebra over $\{a,a^{-1},\mathbb{I}\}$ becomes a division algebra
(other relations concerning division algebras built over the Heisenberg
generators are available in \cite{qsd}).

Let us focus now on the limit $q\to -1$ of the algebra (\ref{qdef}).
This limit can be written as
\begin{eqnarray}
\lim_{q\to -1} a_qa_q^{\dag}-qa_q^{\dag}a_q=
a_{-1}a_{-1}^{\dag}+a_{-1}^{\dag}a_{-1} =:\{a_{-1},a_{-1}^\dag\}=\mathbb{I}.
\end{eqnarray}
A prime remark would be that, recalling that $q\neq 0$, the above limit
could be performed only by avoiding the forbidden value $q=0$;
this can be
done by varying continuously $q$ along a straight line if $q$ does not belong to
the segment $]0,1]$. In the case $q\in ]0,1]$, then the same limit
can be only made by choosing a contour through the complex plane.

Noting that, in any state $|n\rangle$,
\begin{eqnarray}
[N]_{-1}|n\rangle = \lim_{q\to -1} [N]_q |n\rangle = \frac{1-(-1)^N}{2} |n\rangle =
\left\{\begin{array}{cc} 0,& {\rm if}\; n=2p \\
1\,|n\rangle, & {\rm if}\; n=2p+1 \end{array}\right.
\label{limn}
\end{eqnarray}
then we infer the following representation for the operators
\begin{eqnarray}
&\lim_{q\to -1} a_q |n\rangle= \sqrt{\frac{1-(-1)^{N+1}}{2(N+1)}} a |n\rangle
= \sqrt{\frac{1-(-1)^{n}}{2}}  |n-1\rangle =
\left\{\begin{array}{cc}  0 & {\rm if}\; n=2p \\
 1\,|2p\rangle & {\rm if}\; n=2p+1 \end{array}\right.\\
&\lim_{q\to -1} a^\dag_q |n\rangle= \sqrt{\frac{1-(-1)^{N}}{2N}}a^\dag|n\rangle
= \sqrt{\frac{1-(-1)^{n+1}}{2}}  |n+1\rangle =
\left\{\begin{array}{cc} 1\,|2p+1\rangle & {\rm if}\; n=2p \\
 0 & {\rm if}\; n=2p+1 \end{array}\right.
\end{eqnarray}
Let us recall that a fermionic algebra is usually defined by a set of algebraic
relations
\begin{eqnarray}
c c^\dag + c^\dag c = \mathbb{I},\quad c^2 =0= (c^\dag)^2.
\end{eqnarray}
The anticommutation rule is already satisfied by the pair
$(a_{-1}, a_{-1}^\dag)$. Checking, that for any $q\in \mathbb{D}^*$,
$a_q^{2} |n\rangle=\sqrt{[n-1]_q[n]_q}|n-2\rangle$,
$(a_q^\dag)^{2} |n\rangle=\sqrt{[n+1]_q[n+2]_q}|n+2\rangle$,
one infers from (\ref{limn}) that, indeed, for any state, $a_{-1}^{2} |n\rangle=0$
and $(a_{-1}^\dag)^{2} |n\rangle=0$. Thus, the pair $(a_{-1}, a_{-1}^\dag)$
is close to what one usually refers to as a fermionic algebra.
For this reason, we will refer henceforth to these operators to fermions
and to their algebra, to a fermionic algebra.
Here, more rigorously, the operators $(a_{-1}, a_{-1}^\dag)$ are fermionic
operators with an infinite dimensional representation space which is a direct sum
of ordinary two dimensional fermionic representation spaces.

\section{$q$-grading of deformed Heisenberg algebras}

The notion of $\mathbb{Z}_2$ Grassmann grading for associative complex
superalgebras \cite{cornwel} will find, in the next lines, an
extension according to the present $q$-deformed study. But before, for the sake of rigor,
let us put in algebraic terms the definition of the deformation
of the Heisenberg algebra (\ref{qdef}).

\noindent{\bf Building a $q$-grading on the enveloping algebra.}
For all $q\in \mathbb{D}^*$, we introduce
the deformed complex Heisenberg algebra with its  three generators
and deformed commutator as the pair
\begin{eqnarray}
\left(H_q= {\rm span}_{ \mathbb{C}}\{a_q, \,a^\natural_q,\mathbb{I}\}
\;\;;\;\; [(\cdot),(\cdot)]_q\right).
\label{hesdef}
\end{eqnarray}
Some remarks are in order at this stage. First, the deformed Heisenberg algebra $H_q$
is not a Lie algebra unless that one considers the limit points
$q\in \{0,\pm 1\}$. The Jacobi identity fails to be satisfied
 in the general situation when $q\notin \{0,\pm 1\}$. Note also that these algebras
 are not disjoint
since $\mathbb{I}\in H_q \cap H_{q'}$, for $q\neq q'$.
The data of the pair (\ref{hesdef})
are equivalent to the data of a complex vector space $H_q$
and a constraint (equivalence relation) $[a_q,a_q^\natural]_q=\mathbb{I}$
on the tensor algebra built out of its generators.

Next, let us give the definition of the $q$-grading of generators of any
$H_q$ and find an extension for any element of the enveloping algebra
spanned by all $H_q$'s, $q\in \mathbb{D}^*$.
This concept will be introduced by the data of two attributes related
to the parameters $q$:
the ``degree'', denoted by $|(\cdot)|$, and the ``radius'', denoted by $\ell(\cdot)$.

By convention, elements of
$h_0={\rm span}_{ \mathbb{C}}\{\mathbb{I}\}$ are of degree $0$
and we define the degree of the generators of $H_q$ as
\begin{eqnarray}
|a_q|= \sqrt{\frac{\varphi_q}{\pi}}=|a^\natural_q|,\quad
 |\mathbb{I}|=0.
 \label{deg}
\end{eqnarray}
Given a generator of $H_q$, its degree becomes
a real parameter in the segment $[0,2[$ which can be viewed
as the normalized phase of the deformation parameter $q$.
For instance, the degree of an ordinary (Heisenberg) boson
is $|a_{q=1}|= 0=|a^\dag_{q=1}|$, while the degree of the
operators $|a_{q=-1}|=1=|a^\dag_{q=-1}|$ reproducing
a well defined notion of $\mathbb{Z}_2$ Grassmann parity
for these limit.

We will characterize the generators of $H_q$, by another quantity
that we will refer to as its ``radius'' or ``length''
which is nothing but
\begin{eqnarray}
\ell(a_q)=\sqrt{r_q}=\ell(a^\natural_q),\quad
  \ell(\mathbb{I}):=1,
 \label{leng}
\end{eqnarray}
given the modulus $r_q$ of the deformation parameter $q$.

At this stage, the following deformed bracket for elementary generators
can be defined
\begin{equation}
[x_{q},y_{q'}]_{q,q'}:= x_{q}\,y_{q'} -  g(q,q')y_{q'}\,x_{q},\qquad
g(q,q'):=e^{i\pi|x_{q}||y_{q'}|}\ell(x_{q})\ell(y_{q'}).
\label{gradef}
\end{equation}
A quick verification, using (\ref{deg}) and (\ref{leng}), yields the following limits
\begin{eqnarray}
{\rm (boson)}\;\, g(1,1)=1: \;\;[a_{1},a^\dag_{1}]_{1,1} = [a_{1},a^\dag_{1}]_{q=1}=
a_{1} a^\dag_{1} - a^\dag_{1} a_{1} = \mathbb{I},&&
\label{com}\\
{\rm (fermion)}\;\,
g(-1,-1)=-1: \;\,[a_{-1},a^\dag_{-1}]_{-1,-1} = [a_{-1},a^\dag_{-1}]_{q=-1}=
a_{-1} a^\dag_{-1} + a^\dag_{-1} a_{-1} = \mathbb{I},&&
\label{antic}\\
{\rm (q-def.)} \;\,g(q,q)=q: \;\,[a_{q},a^\dag_{q}]_{q,q} = [a_{q},a^\dag_{q}]_{q}=
a_{q} a^\dag_{q} - qa^\dag_{q} a_{q} = \mathbb{I}.&&
\end{eqnarray}
Another interesting property of the deformed bracket (\ref{gradef})
is that it reproduces the $\mathbb{Z}_2$ graded
bracket between fermion and bosons. In other words,
in addition to (\ref{com}) and (\ref{antic}),
the bracket of a fermion and a boson is a commutator,
 because of $g(1,-1)=1$.

Having  properly defined the notion of $q$-grading of basic generators,
let us go further by defining similar ideas for more
complex structures.

The (noncommutative) product of elements of two algebras $H_q$ and $H_{q'}$
lies in the complex vector space $H_q H_{q'}$.
By iteration, one can build monomials in basic generators living in
a product of deformed algebras $H_{q_i}$, $i=1,2,\ldots$. Taking the complex span
of these monomials, one forms a complex vector space. Negative integer powers
of generators can be defined algebraically as $x_q^{-n}:=(x_q^{-1})^n$, $n\in\mathbb{N}$,
where the inverse of $x_q$, i.e. $x_q^{-1}$, acts by representation
such that $ x_q x_q^{-1}=\mathbb{I}$ or $x_q^{-1} x_q=\mathbb{I}$ (right or left
inverse). Further precisions on the division algebra generated
by the Heisenberg operators $a_q$ and $a^\dag_q$ can be found in \cite{qsd}.
The overall algebra spanned by any linear combination of any kind of products
of generators (including inverse integer powers)
will be called the deformed enveloping algebra denoted by ${\mathcal U}_q(H)$.

We would like to give a sense to the notion of grading for any elements of
the enveloping algebra ${\mathcal U}_q(H)$.
We start by the degree and radius of bilinear products
which can be defined as
\begin{equation}
|x_q\, y_{q'}|= |x_q| + \,|y_{q'}|,\qquad \ell(x_q\, y_{q'})= \ell(x_q)\ell(y_{q'}),
\end{equation}
where $x_q$ and $y_{q'}$ are generators of $H_q$ and $H_{q'}$, respectively.
It is remarkable that $|x_q\, y_{q'}|=| y_{q'}\,x_q|$ and $\ell(x_q\, y_{q'})= \ell(y_{q'}\,x_q )$.
Integer powers of elementary generators of $H_q$ belonging to
$(H_q)^\alpha\subset {\mathcal U}_q(H)$ can be also assigned with a degree and a
radius as
\begin{equation}
|(x_q)^{\alpha}|= \alpha |x_q|,\qquad \ell((x_q)^\alpha)= (\ell(x_q))^\alpha,\qquad
\alpha\in\mathbb{Z}.
\end{equation}
More generally, the following relations, valid for finite products of
integer powers of elementary generators, stand for definition:
\begin{eqnarray}
|\prod_{i=1}^n (x_{q_i})^{\alpha_i}|:=
\sum_{i=1}^n \alpha_i|x_{q_i}|,\quad
\ell(\prod_{i=1}^n (x_{q_i})^{\alpha_i}):=
\prod_{i=1}^n (\ell(x_{q_i}))^{\alpha_i}.
\label{prodgen}
\end{eqnarray}
Some products of basic generators admit a spectral decomposition of the form
$A_q =\prod_{i=1}^n x_{q_i}^{\alpha_i}= \sum_{n=0}^{\infty} A_q([n]) |n \rangle \langle n|$,
where $A_q([n])\in \mathbb{C}$, any function of number operators
being a typical example. For this kind of operators,
rational and real powers also  have a rigorous definition.
For instance, one sets $A_q^{\alpha}:= \sum_{n=0}^{\infty}
(A_q([n]))^{\alpha} |n \rangle \langle n|$, $\alpha\in\mathbb{R}$.
In this situation of a possible diagonal decomposition of an operator
being a product of elementary generators, the formulas (\ref{prodgen})
can be extended to real powers:
\begin{eqnarray}
\left|\left(\prod_{i=1}^n (x_{q_i})^{\alpha_i}\right)^{\beta}\right|:=
\beta\sum_{i=1}^n \alpha_i|x_{q_i}|,\quad
\ell(\left(\prod_{i=1}^n (x_{q_i})^{\alpha_i}\right)^{\beta}):=
\prod_{i=1}^n (\ell(x_{q_i}))^{\beta\alpha_i},\quad \beta \in \mathbb{R}.
\label{realpow}
\label{prodgen}
\end{eqnarray}
In order to compute the deformed bracket of composite elements in ${\mathcal U}_q(H)$,
one has to perform first the decomposition in sum of monomials in
the elementary generators before the computation.
Finally, the general bracket for any monomials
$A_{q_i}= \prod_{i_k} x_{q_{i_k}}^{\alpha_{i_k}}$ and
$B_{q_j}= \prod_{j_k} x_{q_{j_k}}^{\alpha_{j_k}}$
can be expressed as follows (an explicit example is provided in the
appendix)
\begin{equation}
[A_{q_i},B_{q_j}]_{G(q_i,q_j)}:= A_{q_i}\,B_{q_j} -
G(q_i,q_j)B_{q_j}\,A_{q_i},
\qquad
G(q_i,q_j):=e^{i\pi|A_{q_i}||B_{q_j}|}\ell(A_{q_i}B_{q_j}).
\label{gradef3}
\end{equation}
In conformity with the above definition of degree,
the degree of a product of two fermions is $2$
(for instance $|a_{-1}a_{-1}^\dag|=1+1=2$)
and not $0$ as it is customarily the case in the context of $\mathbb{Z}_2$
superalgebras. This does not lead to any contradiction
and the anticommutator (\ref{antic}) is still valid and
based on the product of degrees. Although one is tempted
to take for definition of the degree a kind of number modulo $2$
(or $2\pi$ from (\ref{deg})),
the study can be pursued in this general context proving that
there is no need to make further assumptions in the definition (\ref{deg}).

\noindent{\bf $(q=\pm 1)$-grading and matrix representation.}
As a prime interesting feature with implications in supersymmetry,
we discuss the matrix representation.
Let us recall that the ordinary notion of $\mathbb{Z}_2$ grading applied to matrix
algebras \cite{cornwel} can be introduced by the data of two integers $n$ and $m$,
and the decomposition of any element of $M_{n+m}(\mathbb{K})$,
the set of square matrices of dimension $(n+m)^2$ with coefficients
in the field $\mathbb{K}$, into four submatrices of dimensions
$n\times n$, $n\times m$, $m\times n$ and $m\times m$ \cite{cornwel},
i.e.
\begin{eqnarray}
M(n+m,n+m)= \left(\begin{array}{cc}
A(n,n)&B(n,m)\\
C(m,n)&D(m,m)
\end{array}\right).
\end{eqnarray}
A matrix is said to be ``even'' if its entries belong either
to $A(n,n)$ or to $D(m,m)$ and ``odd'' if its entries belong to the
matrices $B(n,m)$ or $C(m,n)$. One can check that any product of matrices
obeys to the law
$even.even=even$, $even.odd=odd$, $odd.odd=even$ such that the matrix
product is stable under this grading.
The same idea can be simply illustrated in ordinary matrix
formulation of supersymmetry where the supersymmetric Hamiltonian
is diagonal (even quantity) and supercharges consist in off diagonal matrices
(odd elements).
One thing remains to be clarified:
if the ordinary bosonic modes $a_1=a$ and
$a_1^\dag=a^\dag$ or usual fermionic operators $c$ and $c^\dag$
admit a matrix representation onto the Fock
Hilbert space basis, it can be suggested to find the equivalent
feature such that the notion of $\mathbb{Z}_2$ grading as previously discussed
can be readily read from their matrix representation.

At first, using the conventional matrix representation
of bosonic modes, nothing can be said. However, if one organizes
the states differently such that we write the Fock basis
in the following form
\begin{eqnarray}
\left\{\left\{ |0\rangle, |2\rangle,|4\rangle,\dots, |2p\rangle,\ldots\right\},
\left\{ |1\rangle, |3\rangle,|5\rangle,\dots, |2p+1\rangle,\ldots\right\}\right\}
\label{orde}
\end{eqnarray}
then the ordinary boson $a_1$ and fermion $a_{-1}$ have the
following matrices with respect to the order (\ref{orde})
{\scriptsize
\begin{equation}
{\normalsize{a_1}} =
\left(\begin{array}{ccccccccc}
0&0&0& \cdots & \cdots & 1 & 0 &  0        & \cdots\\
0&0&0& \cdots & \cdots & 0 & \sqrt{3} & 0 & \cdots\\
0&0&0& \cdots & \cdots & 0 & 0 & \sqrt{5} & \cdots\\
\vdots&\vdots &\vdots& \vdots&\vdots & \vdots &\vdots  &\vdots  &\vdots\\
0&\sqrt{2}&0 & 0 & \cdots&0&0&\cdots&\ldots\\
0&0&\sqrt{4}&0& \cdots&0&0&\cdots&\cdots\\
0&0&0 &\sqrt{6}& \cdots&0&0&\cdots&\cdots\\
\vdots&\vdots &\vdots& \vdots&\vdots & \vdots &\vdots  &\vdots  &\vdots\\
\end{array}\right),\;\;
a_{-1} =
\left(\begin{array}{ccccccccc}
0&0&0& \cdots & \cdots & 1 & 0 &  0        & \cdots\\
0&0&0& \cdots & \cdots & 0 & 1 & 0 & \cdots\\
0&0&0& \cdots & \cdots & 0 & 0 &1 & \cdots\\
\vdots&\vdots &\vdots& \vdots&\vdots & \vdots &\vdots  &\vdots  &\vdots \\
0&0&0 &  \cdots & \cdots&0&0&\cdots&\cdots\\
0&0&0&\cdots& \cdots&0&0&\cdots&\cdots\\
0&0&0 &\cdots& \cdots&0&0&\cdots&\cdots\\
\vdots&\vdots &\vdots& \vdots&\vdots & \vdots &\vdots  &\vdots  &\vdots\\
\end{array}\right).
\label{mat}
\end{equation}
}
The associated adjoint operators can be  easily inferred. In this context, the notion of
``odd'' matrix can be affected to either the pair $(a_{1},a^\dag_{1})$
or to the pair $(a_{-1},a^\dag_{-1})$. Then, legitimately in this context, the
fermions $(a_{-1},a^\dag_{-1})$ can be seen as ``odd'' elements
while the fact that the operators $(a_{1},a^\dag_{1})$ can be
also seen as ``odd'' quantities becomes confusing. Nevertheless,
another interesting feature emerges: multiplying
``odd'' matrices, for instance $a_1 ^\dag a_1$ (of degree $0$),
or $a_{-1}^\dag a_{-1}$ (of degree $2$) will
produce ``even'' elements (as they should be) which are the bosonic and fermionic
numbers, respectively.

\section{Deformed supersymmetry}
\label{sect4}

This section aims at defining a general notion of supersymmetry
on the enveloping algebra ${\mathcal U}_q(H)$. The technical
difficulty comes from the fact that operators
are noncommuting objects in contrast to the situation of ordinary
supersymmetric quantum theory. In addition, for different $q$'s,
all $q$-deformed operators act on an identical Hilbert space (the Fock Hilbert space).

\noindent{\bf $(1,-1)$-Supersymmetry.}
In this paragraph, we define, as a guiding model to next discussions,
a supersymmetric theory on the enveloping subalgebra spanned only
by products in $H_{-1}$ and $H_{1}$. Supersymmetry
is realized through a set of charges commuting with a Hamiltonian.
Operators mapping in a deformed way fermions $a_{-1}$ and $a^\dag_{-1}$ to
bosons $a_{1}$ and $a^\dag_{1}$ are identified. The converse is essentially not
true due to the deformation.

Simple properties allow us to investigate which kind
of operators can generate a supersymmetry. We will restrict the study
to the situation of a supersymmetry generated by only quadratic
products of operators which is, in fact, the closest
possible to the ordinary notion of supersymmetry where
a Hamiltonian appears as a (supersymmetrically) factorized by
bilinear operators: the supercharges \cite{cks}. These latter operators
generically are of the form of a product of bosons and fermions.
These supercharges via a graded structure maps
bosonic to fermionic degrees of freedom and vice versa.

Let us then list the possible minimal bilinears, built from
products of the fermions $a_{-1}$ and $a^\dag_{-1}$ by bosons $a_{1}$ and $a^\dag_{1}$.
Bearing in mind that the order of operators is important,
$6$ monomials are of interest
\begin{eqnarray}
&&
\mathfrak{q}_{1}=a_{-1} a_{1}=\sqrt{\frac{[N+1]_{-1}}{N+1}}\,a^2,\;\;\;\;
\tilde{\mathfrak{q}}_{1}=a_{1}a_{-1}=\sqrt{\frac{[N+2]_{-1}}{N+2}}\,a^2,\\
&&
\tilde{\mathfrak{q}}_{1}^\dag =
a^\dag_{-1}a^\dag_{1}=\sqrt{\frac{[N]_{-1}}{N}}(a^\dag_{1})^2, \;\;\;\;
\mathfrak{q}_{1}^\dag =
a^\dag_{1}a^\dag_{-1}=\sqrt{\frac{[N-1]_{-1}}{N-1}}(a^\dag_{1})^2,\\
&&
\mathfrak{q}_{2}=a^\dag_{1}a_{-1}=\sqrt{\frac{[N]_{-1}}{N}}N=a_{-1}^\dag a_{1},\;\;\;\;
\tilde{\mathfrak{q}_{2}}  =a_{-1}a^\dag_{1}=\sqrt{\frac{[N+1]_{-1}}{N+1}}(N+1)=a_{1}a_{-1}^\dag,
\end{eqnarray}
all of degree\footnote{ By consistency, for instance, we can check that for
$|\mathfrak{q}_{1}|=|a_{-1} a_{1}|= 1 + 0 =
|\sqrt{[N]_{-1}}|+ |a^2| + |N^{-1/2}| =(1/2+1/2)+2\cdot 0+(-1/2)\cdot 0=1$,
obtained from (\ref{realpow}).
The degrees of the other operators can be derived in a similar way.}
 1 and radius $1$.
Note that $\mathfrak{q}_{2}$ and $\tilde{\mathfrak{q}_{2}}$ are self-adjoint.
A set of Hermitian Hamiltonian operators can be readily obtained from
these operators
\begin{eqnarray}
&&
\mathfrak{h}_{1} = [\mathfrak{q}_{1},\mathfrak{q}_{1}^\dag ]_G =
\mathfrak{q}_{1}\mathfrak{q}_{1}^\dag +
\mathfrak{q}_{1}^\dag\mathfrak{q}_{1}= 2[N+1]_{-1}(N+1),\\
&&
\tilde{\mathfrak{h}}_{1} =
[\tilde{\mathfrak{q}}_{1},\tilde{\mathfrak{q}}_{1}^\dag ]_G =
\tilde{\mathfrak{q}}_{1}\tilde{\mathfrak{q}}_{1}^\dag
+ \tilde{\mathfrak{q}}_{1}^\dag\tilde{\mathfrak{q}}_{1}
= 2[N]_{-1}N,
\\
&&
\mathfrak{h}_{2} = (\mathfrak{q}_{2})^2 =[N]_{-1}N,\\
&&
\tilde{\mathfrak{h}}_{2} = (\tilde{\mathfrak{q}_{2}})^2 =[N+1]_{-1}(N+1).
\end{eqnarray}
We are now in position to define the basic Hermitian supersymmetric Hamiltonian
(up to some energy scale $\hbar \omega$ that we omit)
\begin{eqnarray}
\mathfrak{h}_{\rm ss} =  [N]_{-1}N = a^\dag_1 a_1 \, a_{-1}^\dag a_{-1},\;\;\;
|\mathfrak{h}_{\rm ss}|= |[N]_{-1}| + |N| = 2,\quad \ell(\mathfrak{h}_{\rm ss})= 1,
\end{eqnarray}
and the following supersymmetric algebra can be verified
\begin{eqnarray}
&&
[{\mathcal Q}, \mathfrak{h}_{\rm ss}]_{G}=0= [{\mathcal Q}^\dag,\mathfrak{h}_{\rm ss}]_{G},\;\;\;
[{\mathcal Q},{\mathcal Q}^\dag]_G= \mathfrak{h}_{\rm ss},\quad
{\mathcal Q} := \frac{1}{\sqrt{2}}\mathfrak{q}_{1},
\label{susyalg}\\
&&
[\mathfrak{q}_{2},\mathfrak{h}_{\rm ss}]_{G}=0,\quad
\mathfrak{h}_{\rm ss}= (\mathfrak{q}_{2})^2 .
\label{susy2}
\end{eqnarray}
Thus the formulation allows to generate a ${\mathcal N}=3$ supersymmetry
(with three different symmetries).
The other operators $\tilde{\mathfrak{q}}_{1}$ and  $\tilde{\mathfrak{q}_{2}}$
have a simple meaning, in the present context. They define the partner charges
allowing the construction of another supersymmetric Hamiltonian $\tilde{\mathfrak{h}}_{\rm ss}$,
the so-called superpartner of $\mathfrak{h}_{\rm ss}$, obtained by reversing the order of operators,
namely $\tilde{\mathfrak{h}}_{\rm ss}=(\tilde{\mathfrak{q}_{2}})^2 =(1/2)[\tilde{\mathfrak{q}}_{1},
\tilde{\mathfrak{q}}_{1}^\dag]_G$.

To the question ``is this notion of $(1,-1)$-supersymmetry
equivalent to the ordinary one ?'', the answer is no. A hint to
recognize this fact is the form of the ordinary supersymmetric
Hamiltonian $h_{\rm ss}= N_B + N_F$, where $N_B$ and $N_F$ are the
bosonic and fermionic number operators, respectively. Here the
supersymmetric Hamiltonian $\mathfrak{h}_{\rm ss}$ is clearly not of
this form. However, the operator $h_{\rm ss}$ can be rebuilt as
$h_{\rm ss}= \mathfrak{h}_{\rm ss} + \tilde{\mathfrak{h}}_{\rm
ss}=[N]_{-1}N + [N+1]_{-1}(N+1)= N +[N+1]_{-1}$. From this point out
view, the supersymmetries (\ref{susyalg})  or (\ref{susy2}) appear
therefore as the basic ones even though it is not true that all
properties of the ordinary supersymmetry can be recovered. In the
following, we focus on the evidence of the deformation of the
ordinary supersymmetry if the symmetry is realized as
(\ref{susyalg}) or (\ref{susy2}).

Let us check if the ordinary properties of supercharges are satisfied.
First, the square of non Hermitian supercharges are usually vanishing quantities,
here
\begin{eqnarray}
{\mathcal Q}_1^2 \neq  0  \quad  {\rm and } \quad ({\mathcal Q}_1^\dag)^2 \neq 0.
\end{eqnarray}
This shows that the supersymmetry is actually realized in a deformed
way. Second, supercharges used to map bosons onto fermions and
conversely. In the present situation, any $q$-commutation relation
does not lead to interesting results. Nevertheless, after scrupulous
analysis (see Appendix),
 one can reach the following interesting algebras
\begin{eqnarray}
&&[\mathfrak{q}_{1}, a_{-1}^\dag ]_G = [N]_{-1} a_{1}, \label{eq1}\\
&&[\mathfrak{q}_{1}^\dag, a_{-1}^\dag ]_G = [N-1]_{-1} a^\dag_{1}, \label{eq2}\\
&&[\mathfrak{q}_{2}, a_{-1}]_G = [N+1]_{-1} a_{1}, \label{eq3}\\
&&[\mathfrak{q}_{2}, a_{-1}^\dag ]_G = [N]_{-1} a^\dag_{1},
\label{eq4}
\end{eqnarray}
revealing that fermions are actually mapped on deformed bosons
(up to a function of the fermionic number operator).
The converse is not true (see Appendix) pointing out the peculiar
aspects of this deformed supersymmetry.

\noindent{\bf $(q,\bar q)$-Supersymmetry.}
In this last paragraph, we  define in a more general context
of $q$-deformation, the notion of deformed supersymmetry.

We start by the simple remark that, as shown above, elements
of the Heisenberg algebra $H_{q=-1}$ can be seen as deformed
supersymmetric partners of elements of $H_{q=1}$.
The complex number $-1$ is obtained after a rotation by $\pi$
from the complex number $1$. The notion of supersymmetry could
find an equivalence in ${\mathcal U}_q(H)$, mapping
generators of $H_q$ onto a generator of $H_{\bar q}$, with
$\bar q$ a transformation of $q$:
\begin{eqnarray}
\bar q=S_{f,k}(q)= f(r_q,\varphi) e^{ik(r_q,\varphi)},\qquad
0< f(r_q,\varphi)\leq 1,\quad 0\leq k(r_q,\varphi) < 2\pi,
\end{eqnarray}
where $f$ and $k$ are real functions.
The above case of $(1,-1)$-supersymmetry corresponds to
$S_{Id,\pi}$, a simple rotation $R_\pi(\cdot)=e^{i\pi}(\cdot)$
in the complex plane by an angle $\pi$.
However, we will assume that $f\equiv f(r_q)$ and $k\equiv k(\varphi_q)$;
so doing, we write $\bar q = f(r_q)e^{i k(\varphi_q)}$, thereby defining
$r_{\bar q}:= f(r_q)$ and $\varphi_{\bar q}:=k(\varphi_q)$.

The second step is to define again bilinears which are
of interest. It does not take long to find the following operators
(the same notation as in the previous paragraph is used but operators
now refer to different quantities)
\begin{eqnarray}
&&
\mathfrak{q}_{1}=a_{\bar q} a_{q}=
\sqrt{\frac{[N+1]_{\bar q}[N+2]_{q}}{(N+1)(N+2)}}\,(a_1)^2,\;\;\;\;
\tilde{\mathfrak{q}}_{1}= a_{q} a_{\bar q}=
\sqrt{\frac{[N+1]_{q}[N+2]_{\bar q}}{(N+1)(N+2)}}\,(a_1)^2,\\
&&
\tilde{\mathfrak{q}}_{1}^\natural =
a^\natural_{\bar q}a^\natural_{q}=
\sqrt{\frac{[N]_{\bar q}[N-1]_{q}}{N(N-1)}}(a^\dag_{1})^2, \;\;\;\;
\mathfrak{q}_{1}^\natural =
a^\natural_{q}a^\natural_{\bar q}=
\sqrt{\frac{[N]_q[N-1]_{\bar q}}{N(N-1)}}(a^\dag_{1})^2,\\
&&
\mathfrak{q}_{2}=a^\natural_{q}a_{\bar q}=\sqrt{[N]_{\bar q}[N]_q}=
a_{\bar q}^\natural a_{q},\;\;\;\;
\tilde{\mathfrak{q}_{2}}  =
a_{\bar q}a^\natural_{q}=
\sqrt{[N+1]_{\bar q}[N+1]_q}= a_{q}a_{\bar q}^\natural.
\end{eqnarray}
They all share the same degree $\sqrt{\varphi_{q}/\pi} + \sqrt{\varphi_{\bar q}/\pi}$
and the same radius $\sqrt{r_q f(r_q)}$.
We also notice that $\mathfrak{q}_{2}$ and $\tilde{\mathfrak{q}_{2}}$ remain Hermitian,
while $\mathfrak{q}_{1}$  and $\tilde{\mathfrak{q}}_{1}^\natural$ become adjoint
of one another.

In order to obtain a Hermitian Hamiltonian operator,
we consider the operator generated by $\mathfrak{q}_{2}$
\begin{eqnarray}
\mathfrak{h}_{2} = (\mathfrak{q}_{2})^2 =[N]_{\bar q}[N]_q=(\{N\}_{\bar q,q})^2,\qquad
|\mathfrak{h}_{2}|= 2\left(\sqrt{\frac{\varphi_{q}}{\pi}} +
\sqrt{\frac{\varphi_{\bar q}}{\pi}}\right), \quad \ell(\mathfrak{h}_2) =  r_q f(r_q)
\label{ham}
\end{eqnarray}
and require that $\mathfrak{q}_{2}$ is a symmetry of $\mathfrak{h}_{2}$
with respect to the deformed bracket. We are led to
the following
\begin{eqnarray}
[\mathfrak{q}_{2}, \mathfrak{h}_{2}]_G=
\sqrt{[N]_{\bar q}[N]_q}[N]_{\bar q}[N]_q \left[1
-e^{ 2i\pi\left(\sqrt{\frac{\varphi_{q}}{\pi}}
+ \sqrt{\frac{\varphi_{\bar q}}{\pi}}\right)^2}(r_q f(r_q))^{\frac{3}{2}}\right]
\end{eqnarray}
A set of necessary and sufficient conditions for
$[\mathfrak{q}_{2}, \mathfrak{h}_{2}]_G= 0$ to hold is
\begin{eqnarray}
\left(\sqrt{\frac{\varphi_{q}}{\pi}} + \sqrt{\frac{\varphi_{\bar q}}{\pi}}\right)^2
= p\in \mathbb{N},\qquad r_q f(r_q) =1,
\end{eqnarray}
which can be easily solved by
\begin{eqnarray}
&&p=0,\;\; \varphi_q=0\;\; {\rm and} \;\; \varphi_{\bar q}=k_0(\varphi_q) =0;
\qquad  f(r_q) =\frac{1}{r_q};\\
&&
p\geq 1,\;\; k_p(\varphi_q) = \varphi_{\bar q}=
(\sqrt{\varphi_{q}}- \sqrt{p\pi})^2 \in [0,2\pi[;
\qquad f(r_q) =\frac{1}{r_q}.
\label{phas}
\end{eqnarray}

\begin{figure}
 \centering
       \begin{minipage}[t]{0.45\textwidth}
      \centering
\includegraphics[angle=0, width=6cm, height=6cm]{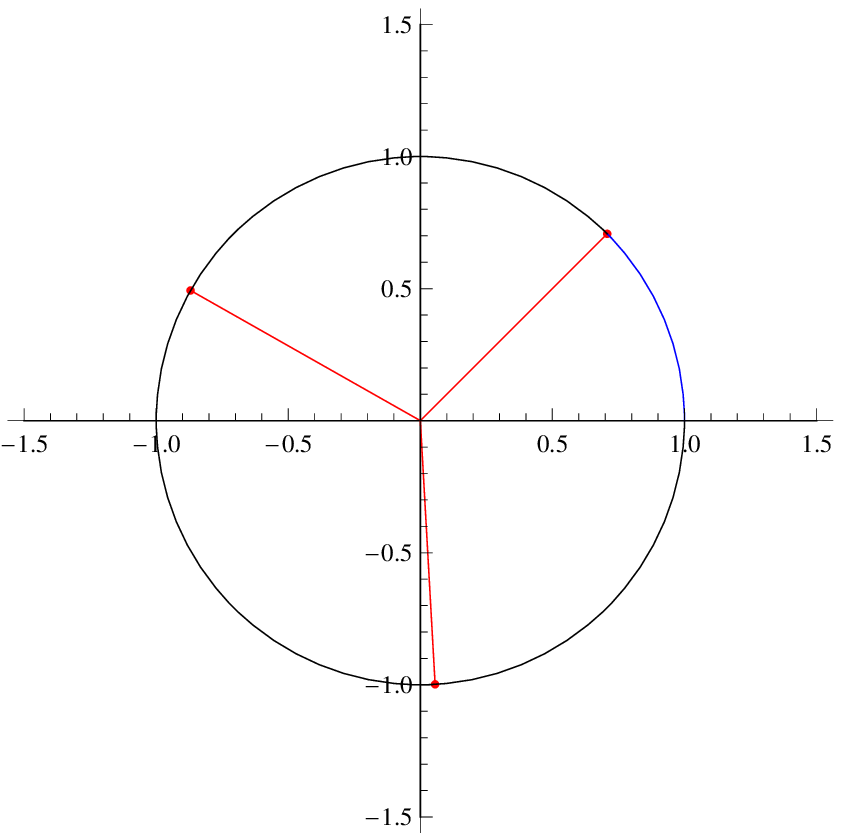}
\caption{
{\small The two possible deformation parameters in $[0,2\pi[$ corresponding
to $q= e^{i\frac{\pi}{4}}$: ${\bar q}_{2}= $ $e^{i\pi(\sqrt{2}-\frac{1}{2})^2}$;
 ${\bar q}_{3}= e^{i\pi(\sqrt{3}-\frac{1}{2})^2}$.
}}
\end{minipage}
\hspace{0.5cm}
 \put(-60,125){ {\small $q=e^{i\frac{\pi}{4}}$} }
  \put(-165,115){ {\small ${\bar q}_1$} }
 \put(-100,20){ {\small ${\bar q}_2$} }
  \put(123,45){ {\small $q$}}
   \put(150,82){ {\small $q_3$}}
   \put(158,65){ {\small $q_1$}}
   \put(160,48){ {\small $q_2$}}
   \put(126,100){ {\small $q_4$}}
   \put(70,85){ {\small $q_5$}}
   \put(62,27){ {\small $q_6$}}
   \put(143,20){ {\small $q_7$}}
            \put(120,175){ {\small $1/x$}}
     \hspace{0.5cm}
      \begin{minipage}[t]{0.45\textwidth}
      \centering
\includegraphics[angle=0, width=4.3cm , height=6.5 cm]{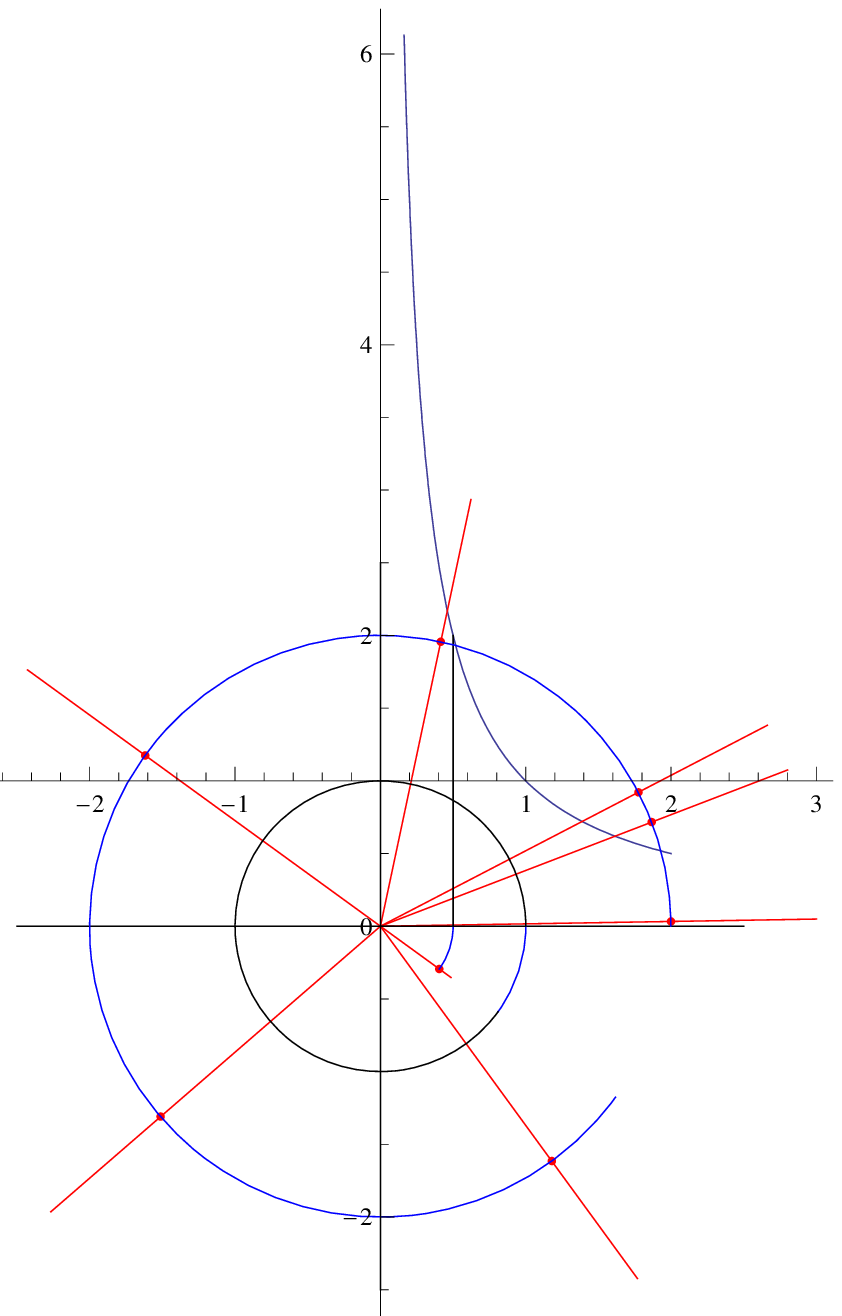}
\caption{{\small Construction of seven possible deformation parameters $\bar q$
lying outside of the unit disc corresponding to $q= \frac{1}{2}e^{i\frac{9\pi}{5}}$:
${\bar q}_{p}= 2e^{i\pi(\sqrt{p}-\sqrt{\frac{9}{5}})^2}$,
$p=1,2,\dots,7$.}}
 \end{minipage}
\end{figure}

Before regarding the phase $\varphi_q$ problems, let us focus on
the equation $f(r_q) =1/r_q$ having a drastic consequence. Indeed,
if $r_q \in ]0,1]$, then $f(r_q)\geq 1$.
This is to say that if we impose $\bar q \in \mathbb{D}^*$,
then obligatory we will restrict to the situation where $r_q=f(r_q)=1$ in order
to get relevant solutions.
Enlarging the scope of value\footnote{
From the beginning, we never assume this condition which is without
any consequence on the present formulation.} of $q\in \mathbb{C}$, another
interesting feature emerges: superpartners of operators
with deformation parameter $q$ strictly inside of the unit disc $\mathbb{D}^*$
are operators with parameter  $\bar q$ lying strictly outside of the unit disc
(and vice-versa);
superpartners of operators labeled by a $q$ belonging to the unit circle
$S^1= \{|z|=1, z\in \mathbb{C}\}$ are operators with parameter
still on the circle.

Concerning the phase, the case $p=0, \varphi_q=0=k_0(\varphi_q)$,
for $r_q=1=f(r_q)$ refers to the trivial point $q=1=\bar q$.
We will focus only on points ${\bar q}\neq q$ encoding
values where partners differ from one another.
It turns out that for $\varphi_q \in \{0,\pi/4,\pi/2, 3\pi/4,\ldots, 7\pi/4\}$,
there always  exists a value ${\bar q} = q$ since one can show that
$p=(4/\pi)\varphi_q\in [\![ 0,7]\!]$.
More generally, solving the first condition of (\ref{phas}),
for $p\geq 1$, one infers the following constraints on $\varphi_q$
\begin{eqnarray}
&&
\forall \varphi_q\in [0,2\pi [,\;\;\;p=1,2,\;\;
k_p(\varphi_q) = \left(\sqrt{p\pi} - \sqrt{\varphi_{q}}\right)^2 \in [0,2\pi[ ,\label{k+}\\
&&
\forall\varphi_{q}\in [\pi \left(\sqrt{2}-\sqrt{p}\right)^2,2\pi[,\;\;\;p\in [\![ 3,7]\!],\;\;
k_p(\varphi_q) = \left(\sqrt{p\pi} - \sqrt{\varphi_{q}}\right)^2 \in [0,2\pi[.
\end{eqnarray}
These solutions imply that, regardless of their modulus which can be fixed
to be equal $r_q=1$, given $q$ fixing once for all $r_q$ and $\varphi_q$,
there are at least $2$ and at most $7$ parameters ${\bar q}_p$,
$p\in [\![ 1,7]\!]$, providing good parameter candidates for
different supersymmetries $\mathfrak{q}_{2; p}$ for at least $2$
and at most $7$ different models defined by $\mathfrak{h}_{2; p}$,
such that $[\mathfrak{q}_{2; p},\mathfrak{h}_{2; p}]_G=0$
(see Figure 1 and Figure 2).
In the particular situation of the so-called $(1,-1)$-supersymmetry
as built in the previous paragraph,
given $(q=1, r_1=1,\varphi_{1}=0)$,
the above solutions are consistent and reduce to the unique
possibility of $p=1$ in (\ref{k+}) such that
$k_1(0)=\pi$, thus implying a unique choice for ${\bar q}=e^{i\pi}$.
Finally, the constraint such that $k(\varphi_q)\in [0,2\pi[$ is actually
too strong and more solutions can be determined by relaxing this condition.
In fact, it seems that an infinite set of inequivalent (modulo $2\pi$) solutions
of the phase equation (\ref{phas}) are available due to the fact that
$k(\varphi_q)$ is a nonlinear function of $\varphi_q$.
A careful analysis and the meaning of these solutions
will be treated in a subsequent work.

The meaning of supersymmetry is not clear when considering the other
operators $\mathfrak{q}_{1}$ and $\tilde{\mathfrak{q}}_{1}^\natural$.
A  Hermitian Hamiltonian can be easily identified
using the deformed bracket $\mathfrak{h}_{1}=[\mathfrak{q}_{1},\tilde{\mathfrak{q}}_{1}^\natural]_G$.
However, the set of conditions in order to impose
$[\mathfrak{q}_{1},\mathfrak{h}_{1}]= 0=[\tilde{\mathfrak{q}}_{1}^\natural,\mathfrak{h}_{1}]$
is more involved. Moreover, the resulting operator
$\mathfrak{h}_{1}$ is not equal to $\mathfrak{h}_{2}$ (\ref{ham}).
Other more complicated issues arise when one reverses
the order of the basic generators. The former
${\mathcal N}=3$ $(1,-1)$-deformed supersymmetry
is therefore explicitly broken in the general deformation theory,
with a reduced  number of supercharges equal to ${\mathcal N}=1$.

Let us turn to the properties of the mapping boson-fermions.
 The following relations can be obtained:
\begin{eqnarray}
&&[\mathfrak{q}_{2}, a_{\bar q}]=
{\textstyle\left[\sqrt{\frac{[N]_{\bar q}[N+1]_{\bar q}[N]_q}{[N+1]_q}}-
e^{i\sqrt{\varphi_{\bar q}(\varphi_{\bar q} + \varphi_{q})}}
f(r_q)\sqrt{r_q}[N+1]_{\bar q}\right] a_q},\label{gende}\\
&&[\mathfrak{q}_{2}, a_{\bar q}^\natural]=
{\textstyle\left[[N]_{\bar q}-
e^{i\sqrt{\varphi_{\bar q}(\varphi_{\bar q} + \varphi_{q})}}
f(r_q)\sqrt{r_q}\sqrt{\frac{[N-1]_{\bar q}[N]_{\bar q}[N-1]_q}{[N]_q}}
\right] a^\natural_q},\label{gende2},
\end{eqnarray}
which at the limit $q=1$ and ${\bar q}=-1$, characterized by
$f(r_{1})=1=r_1$, $\varphi_0=0$ and $\varphi_{-1}=\pi$,
reproduce correctly (\ref{eq3}) and (\ref{eq4}).
Hence, one draws the conclusion that  $a_{\bar q}$ and $a^\natural_{\bar q}$
are deformed partners of $a_{q}$ and $a^\natural_{q}$, respectively.

\section{Conclusion}
\label{ccl}

We have succeeded in setting a notion of $q$-grading onto the deformed
enveloping algebra built from all possible products
of $q$-deformed Heisenberg algebras
for different parameters $q\in \mathbb{D}^*$, $\mathbb{D}^*$ being
the complex disc of radius one without $0$.
This notion of $q$-grading encompasses in
specific limit the ordinary notion of $\mathbb{Z}_2$ grading
of ordinary associative superalgebra. A generalized bracket
is defined on the enveloping algebra which reproduces
the ordinary bosonic, fermionic,
$\mathbb{Z}_2$ graded and $q$-deformed commutators
for corresponding subspaces in  the total algebra.
The formalism is then used to show
that the notion of supersymmetry can be extended and, even,
realized in the present situation where the fermions do not commute with bosons.
In the specific instance such that $q=\pm 1$, a supersymmetric
Hamiltonian has been defined and its ${\mathcal N}=3$ supersymmetry properly identified
with new deformed features. Finally, in the full $q$-deformed theory,
we have identified many kinds of operators
(restricting the deformed phase parameter in $[0,2\pi[$)
which are able to define different supersymmetric models.
In the general context of deformation, the  ${\mathcal N}=3$
$(1,-1)$-supersymmetry is explicitly broken and at least ${\mathcal N}=1$
supercharge can be defined in any $(q,{\bar q})$-deformed supersymmetric model.

\section*{Acknowledgments}
M.N.H. thanks the National Institute for Theoretical Physics (NITheP) and its Director
Prof. Frederik G. Scholtz for hospitality during a pleasant stay in Stellenbosch where
this work has been initiated.
J.B.G. thanks Pr. Micha\"el Kastner, Pr. Jan Govaerts and Pr. Bo-Sture Skargerstam for
helpful discussions. This work was supported under a grant
of the National Research Foundation of South Africa
and by the ICTP through the OEA-ICMPA-Prj-15.

\section*{Appendix}
\label{App}

This appendix provides useful relations and explicit illustrations to the text.

We start by illustrating the kind of computations involved by the generalized
$q$-graded bracket. Let us calculate, for the specific instance,
the bracket of the monomials of the form $a_{q_1}^\natural a_{q_1'}$ and
$a_{q_2}^\natural a_{q_2'}a_{q_3}^\natural a_{q_3'}$
(as appeared in  computing  the bracket of the supercharge
$\mathfrak{q}_{2}=a^\dag_{1}a_{-1}$ and  supersymmetric
  Hamiltonian $\mathfrak{h}_{\rm ss}=a^\dag_1 a_1 \, a_{-1}^\dag a_{-1}$).
Given the complex numbers $q_i=r_{q_i}e^{i\varphi_{q_i}}$
and  $q'_i=r_{q'_i}e^{i\varphi_{q'_i}}$, $i=1,2,3$, we have
\begin{eqnarray}
&& |a_{q_1}^\natural a_{q_1'}|=\sqrt{\frac{\varphi_{q_1}}{\pi}}+
\sqrt{\frac{\varphi_{q'_1}}{\pi}},\quad
|a_{q_2}^\natural a_{q_2'}a_{q_3}^\natural a_{q_3'}|
=\sqrt{\frac{\varphi_{q_2}}{\pi}}+ \sqrt{\frac{\varphi_{q'_2}}{\pi}}
+\sqrt{\frac{\varphi_{q_3}}{\pi}}+ \sqrt{\frac{\varphi_{q'_3}}{\pi}}\cr
&&\ell(a_{q_1}^\natural a_{q_1'}) = \sqrt{r_{q_1}r_{q_1'} }, \quad
\ell(a_{q_2}^\natural a_{q_2'}a_{q_3}^\natural a_{q_3'})
= \sqrt{r_{q_2}r_{q_2'}r_{q_3}r_{q_3'}},\cr\cr
&&[a_{q_1}^\natural a_{q_1'}, a_{q_2}^\natural a_{q_2'}a_{q_3}^\natural a_{q_3'}]_G
=
a_{q_1}^\natural a_{q_1'} a_{q_2}^\natural a_{q_2'} a_{q_3}^\natural a_{q_3'}\cr
&&
-
(-1)^{|a_{q_1}^\natural a_{q_1'}||a_{q_2}^\natural a_{q_2'}a_{q_3}^\natural a_{q_3'}|}
 \sqrt{r_{q_1}r_{q_1'}r_{q_2}r_{q_2'}r_{q_3}r_{q_3'}}
a_{q_2}^\natural a_{q_2'}a_{q_3}^\natural a_{q_3'} a_{q_1}^\natural a_{q_1'}\cr
&&=
a_{q_1}^\natural a_{q_1'} a_{q_2}^\natural a_{q_2'} a_{q_3}^\natural a_{q_3'}
\label{cal}\\
&&-
 (-1)^{(|a_{q_1}|+|a_{q_1'}^\dag|)(|a_{q_2}^\natural|+| a_{q_2'}|+|a_{q_3}^\natural|+| a_{q_3'}|)}
 \sqrt{r_{q_1}r_{q_1'}r_{q_2}r_{q_2'}r_{q_3}r_{q_3'}}
a_{q_2}^\natural a_{q_2'}a_{q_3}^\natural a_{q_3'}a_{q_1}^\natural a_{q_1'} .\nonumber
\end{eqnarray}
Now fixing $q_1=1$, $q_1'=-1$, $q_2=1=q_2'$ and $q_3=-1=q_3'$,
the expression (\ref{cal}) reduces to
\begin{eqnarray}
[\mathfrak{q}_{2},\mathfrak{h}_{\rm ss}]&=&
[a_{1}^\dag a_{-1}, a^\dag_1 a_1 \, a_{-1}^\dag a_{-1}]_G\cr
&=&
a_{1}^\dag a_{-1}a^\dag_1 a_1 \, a_{-1}^\dag a_{-1}
- (-1)^{(1+0)\cdot(0+0+1+1)}
 a^\dag_1 a_1 \, a_{-1}^\dag a_{-1} a_{1}^\dag a_{-1}\cr
&=&
\sqrt{[N]_{-1}N} [N]_{-1}N
- [N]_{-1}N \sqrt{[N]_{-1}N} =0,
\label{cal2}
\end{eqnarray}
showing that $\mathfrak{q}_{2}$ is a symmetry of $\mathfrak{h}_{\rm ss}$.
Of course, using an ordinary commutator this statement becomes obvious
since both $\mathfrak{q}_{2}$ and $\mathfrak{h}_{\rm ss}$ are pure functions of
the number operator $N$. But here, the difficulty resides in the
fact that we use a different commutation relation which turns out
to simplify in the form of the ordinary commutator.

We give the complete set of commutation relations
between the supercharges $\mathfrak{q}_{1}$, $\mathfrak{q}_{1}^\dag$
and $\mathfrak{q}_{2}$ and the basic
degrees of freedom $a_1$, $a^\dag_1$, $ a_{-1}$ and $ a_{-1}^\dag$.
The following relations hold:

$\bullet$ $\mathfrak{q}_{1} = a_{-1}a_1$
\begin{eqnarray}
&&
[\mathfrak{q}_{1}, a^\dag_{1}]_G = (N+2)a_{-1} - \sqrt{[N]_{-1}N}a_1=
{\textstyle {\left[  (N+2)\sqrt{\frac{[N+1]_{-1}}{N+1}} - \sqrt{[N]_{-1}N} \right]}}a_1,
\label{equa1}\\
&&
[\mathfrak{q}_{1}, a^\dag_{-1}]_G = [N]_{-1} a_{-1},\\
&&
[\mathfrak{q}_{1}, a_{1}]_G =
{\textstyle {\left[\sqrt{\frac{[N+1]_{-1}}{N+1}}-\sqrt{\frac{[N+2]_{-1}}{N+2}}\right]}} (a_{1})^3,\\
&&
[\mathfrak{q}_{1}, a_{-1}]_G =
{\textstyle {\sqrt{\frac{[N+1]_{-1}[N+3]_{-1}}{(N+1)(N+3)}}(a_{1})^3=
\frac{[N+1]_{-1}}{\sqrt{(N+1)(N+3)}}(a_{1})^3}}.
\label{equa2}
\end{eqnarray}

$\bullet$ $\mathfrak{q}^\dag_{1} = a^\dag_{1}a^\dag_{-1}$
\begin{eqnarray}
&&
[\mathfrak{q}^\dag_{1}, a^\dag_{1}]_G =
{\textstyle {
\left[ \sqrt{\frac{[N-1]_{-1}}{N-1}} - \sqrt{\frac{[N-2]_{-1}}{N-2}} \right]}}(a^\dag_1)^3, \\
&&
[\mathfrak{q}^\dag_{1}, a^\dag_{-1}]_G =
{\textstyle {\frac{[N]_{-1}}{\sqrt{N(N-2)}}}}(a^\dag_{1})^3,\\
&&
[\mathfrak{q}^\dag_{1}, a_{1}]_G = a^\dag_{1} \sqrt{[N]_{-1}N} - (N+1) a^\dag_{-1}=
{\textstyle {\left[ \sqrt{[N-1]_{-1}(N-1)} - (N+1)\sqrt{\frac{[N]_{-1}}{N}}\right] }}
a^\dag_{1},\\
&&
[\mathfrak{q}^\dag_{1}, a_{-1}]_G = [N-1]_{-1} a^\dag_{1}.
\end{eqnarray}

$\bullet$ $\mathfrak{q}_{2} = a^\dag_{1}a_{-1}$
\begin{eqnarray}
&&
[\mathfrak{q}_{2}, a^\dag_{1}]_G =
\left[ \sqrt{[N]_{-1}N} - \sqrt{[N-1]_{-1}(N-1)} \right]a^\dag_1, \\
&&
[\mathfrak{q}_{2}, a^\dag_{-1}]_G =[N]_{-1}a^\dag_{1},\\
&&
[\mathfrak{q}_{2}, a_{1}]_G = \sqrt{[N]_{-1}N}a_{1}  - (N+1) a_{-1}=
\left[ \sqrt{[N]_{-1}N} - \sqrt{[N+1]_{-1}(N+1)}\right] a_{1},\\
&&
[\mathfrak{q}_{2}, a_{-1}]_G = \sqrt{[N+1]_{-1}(N+1)} a_{-1}=[N+1]_{-1} a_{1}.
\end{eqnarray}
In summary, these operators map bosons on a mixed combination of
bosons and fermions which cannot be reduced to a deformation of
a fermion. On the other hand, fermions can be exactly mapped onto
bosons up to a deformation function or a nonlinearity
appearing as a power of a bosonic operator.

\end{document}